\documentstyle[12pt,epsfig,amstext]{article}

\begin{document}

\title{ Dimensional crossover  in fragmentation }

\author{ Oscar Sotolongo-Costa$^{1,2}$\footnote{Oscar.Sotolongo@brunel.ac.uk, oscarso@ff.oc.uh.cu},
 Arezky H. Rodriguez$^1$\footnote{arezky@ff.oc.uh.cu}
 and G.J.Rodgers$^2$\footnote{g.j.rodgers@brunel.ac.uk}}

\maketitle
\begin{center}
\footnotesize{1.- Department of Theoretical Physics, Havana University, Habana 10400, Cuba.\\
 2.- Department of Mathematical Sciences, Brunel University,\\
Uxbridge, Middlesex UB8 3PH, UK.\\}
\end{center}
\begin{abstract}
Experiments in which thick clay plates and glass rods are
fractured have revealed different behavior of fragment mass
distribution function in the small and large fragment regions. In
this paper we explain this behavior using non-extensive Tsallis
statistics and show how the crossover between the two regions is
caused by the change in the fragments' dimensionality during the
fracture process: We obtain a physical criterion
 for the position of this crossover and an expression for the change in the power law exponent between the small and large
 fragment regions. These predictions are in good agreement with the experiments on thick clay plates.
\end{abstract}

\noindent PACS number(s): 05.40.Fb, 24.60.-k

\newpage

Experiments on plate fragmentation \cite{1} have revealed two
regions of  behavior in the fragment size distribution
 function (FSDF). The crossover between the regions
  scales with the width of the plate. In these experiments (and also in \cite{2} where an analogous behavior
  was observed in the breaking of glass rods)
   the transition from one kind of behavior to another is assigned to a dimensional crossover, as was also pointed
   out in \cite{3}.
It seems physically clear that a dimensional dependence exists in
the fragmentation of objects. A long
 thin glass rod must reveal a different behavior for fracture than a ``thick cylinder'' of the same material,
 as the small fragments produced from the fracture of glass rods
  appear to be approximately cylindrical. Also, a plate should manifest a
  different behavior for fracture than a ``three-dimensional'' fragment for which all dimensions are
  similar.\\
  In \cite{3} it was shown that no crossover was present for fragmentation of three-dimensional objects. This
   was confirmed in \cite{4} where experiments with falling mercury drops, which always preserve their spherical
    shape, were reported.\\
    Dimensional dependence and multiscaling in fragmentation have
    received attention in \cite{5,6} with {\it{ad hoc}} models built  to describe the dimensional
    dependence of FSDF  and its multifractality.\\
On the other hand attempts have been made to obtain the FSDF starting
from first principles, i.e., the maximum entropy principle
\cite{7,8}. As  the maximum entropy principle is universal, it has
an almost unlimited range of applications, and consequently some of
the properties of FSDF observed in experiments, such as scaling
and multifractality,  are expected to be  revealed. However, no dimensional crossover was found in either \cite{7} or \cite{8}. In our opinion this is due to
the assumption that not just the maximum entropy principle, but
the formula for the Boltzmann-Gibbs (BG) entropy of a system, are
valid for the description of fracture phenomena. The formula for
the BG entropy is: $$S=-k\sum_{i=1}^{W}p_{i}\log p_{i} ,$$\\
 where $p_{i}$ is the
probability of finding the system in the microscopic state $i$,
$k$ is Boltzmann's constant and $W$ is the total number of
microstates.\\ This formula has been shown to be restricted to the
domain of validity of BG statistics, which seem to describe nature
when the effective microscopic interactions and the microscopic
memory are short ranged \cite{9}. The process of shock
fragmentation, specially when energies are high enough, leads to
the existence of long range correlations between all parts of the
object being fragmented. Then the use of the above formula to
describe fragmentation processes seems to be inadequate. From
here it can be concluded that the use of  statistics able to
describe long-range and long-memory interactions can be useful in
the description of fracture
processes.\\
 In this work we report the deduction
of the FSDF and its dimensional crossover from the maximum entropy
principle using as a starting point Tsallis entropy \cite{9}:

\begin{equation}
S_q=k \frac{1-\int_0^\infty p^q(x)dx }{q-1}.  \label{eq:1}
\end{equation}

The integral runs over all admissible values of the magnitude $x$
and $p(x)dx$  is the probability of the system being in a state
between $x$ and $x+dx$.$k$ is the Boltzmann constant and $q$ a
real number (the entropic index).The Tsallis entropy $S_{q}$
reduces to the BG entropy for $q=1$ so recovering Boltzmann-Gibbs
statistics. \\
Dealing with the case of fracture $x$ could well describe the
mass (volume) of the fragments. The statistics based on this
entropy have been used to describe a number of non-equilibrium
processes and phenomena for which BG statistics is not
appropriate. (see \cite{9}), although it seems to be still far
from having shown all its capabilities.
\\

Let us extremize $\frac{S_{q}}{k}$ with appropriate constraints. If
we denote the volume of a fragment by $V$ and some typical volume
characteristic of the distribution by $V_m$, we can define a
dimensionless volume $v=\frac{V}{V_m} $. The normalization
condition reads
\begin{equation}
\int_0 ^\infty p(v)dv =1 .  \label{eq:2}
\end{equation}

The other condition to be imposed is mass conservation. But as the
system is finite, this condition will lead to a very sharp decay
in the asymptotic behavior of the fragment size distribution
function (FSDF) for large sizes of the fragments.
 Consequently, we will impose a more general condition,
like the ``q-conservation'' of the mass, in the form:

\begin{equation}
\int_0^\infty vp^q(v)dv=1,   \label{eq:3}
\end{equation}

which reduces to the ``classical'' mass conservation when $q=1$.

Using the method of Lagrange multipliers we construct the
function:

\begin{equation}
L(p;\alpha_1;\alpha_2)= S_q- \alpha_1 \int_0^\infty p(v)dv
+\alpha_2 \int_0^\infty p^q(v)vdv \label{eq:4}
\end{equation}
The Lagrange multipliers  $\alpha_1$ and $\alpha_2$  are
determined by the constraints. The extremization of $L(p;
\alpha_1; \alpha_2)$ leads to:
\begin{equation}
p(v)dv= C(1+(q-1)\alpha_2v)^{-\frac{1}{q-1}}dv \label{eq:5}
\end{equation}
where the constant $C$ is given by $$C={[\frac{q-1}{q} \alpha_1]}
^{\frac{1}{q-1}}.$$\\ This is a FSDF expressed as a function of
the volume of the fragments. It is valid for $1<q<2 $. The FSDF
given by Eq. \ref{eq:5} can describe satisfactorily the behavior
for ``small'' and ``large'' fragment sizes. Indeed, we can apply
this
equation to the fragmentation of plates described in \cite{1}.\\
As the object to be broken has mainly a two-dimensional shape,
large fragments show a mass scaling with the surface of the basis
of the plate, so that if the width of the plate in units of a
characteristic length of the system is $\Delta$ then the mass
(volume) scales with the linear dimensions as $v \sim \Delta
l^{2}$. This must be taken into account when calculating the
element of volume $dv$. Then, for large fragments we have

\begin{equation}
p_+(l)dl=C l[1+(q-1)\alpha_2l^2\Delta)]^{-\frac{1}{q-1}}dl.
\label{eq:7}
\end{equation}

The small fragments do not resemble plates but volumetric objects.
This means that the volume of the fragment scales as $l^3$ and we
have for the distribution of small fragments:
\begin{equation}
p_-(l)dl=Cl^2[1+(q-1)\alpha_2l^3]^{-\frac{1}{q-1}}dl, \label{eq:8}
\end{equation}

From \ref{eq:7} and \ref{eq:8} we may deduce the asymptotic
behavior to obtain the slope $\beta$ for small and large
fragments, with the notation that the distribution function has
the asymptotic behavior $p(l)\sim l^{-\beta} $ as in \cite{1}.
Designating the slope for large (small) fragments by
$\beta_+$($\beta_-$) respectively, we obtain:

\begin{equation}
\frac{\beta_+}{\beta_-}=\frac{3-q}{5-2q}, \label{eq:9}
\end{equation}
from where the values of $\frac{\beta_+}{\beta_-}$ can be
calculated restricting  $q$ to its range of validity $1<q<2$. This ratio
lies in the interval $\frac{2}{3}<\frac{\beta_+}{\beta_-}<1$.\\
These values should be regarded as a coarse grained estimate,
since to obtain them we have postulated a very definite scaling of
the mass with the dimensionality, although it is clear that this
dependence should be no more than approximate, since the objects
are not exactly ``one'' or ``two-dimensional''. Yet it will be
seen that this simple model is good enough to describe the
behavior of the FSDF.\\ From \cite{1} we may obtain that all the
values of the ratio $\frac{\beta_+}{\beta_-} $ in the reported
experiment satisfy this condition, going from .67 to .79. This is
illustrated in table 1\\
\\

\hspace{100pt}
\begin{tabular}{|c|c|c|c|}
  \hline
   plate  &$\beta_-$&$\beta_+$ &$\frac{\beta_+}{\beta_-}$  \\
  \hline
  2 & 1.62 & 1.19 & 0.73 \\
  3 & 1.5 &1.17  & 0.78 \\
  4 & 1.67 & 1.12 & 0.67 \\
  5 & 1.5 & 1.9 & 0.79 \\ \hline
\end{tabular}
\bigskip
\bigskip
\\
Following the same reasoning we could predict the value of
$\frac{\beta_+}{\beta_-}$ for the experiments reported in
\cite{2}. In this case, as reported there, it is the cumulative
number which has slope $\beta$. In that case, respecting the
notation in \cite{2}, the behavior of the distribution function
$p(l)$ should be $p(l)\sim l^{-\beta-1}$. We predict that for the
breaking of rods, where the crossover is from one-dimensional to
three-dimensional objects the value of the ratio should be around
$3$ irrespective of $q$. New experimental results in this case
would be very welcome to investigate our predictions. \\ From the
present formulation we can evaluate the order of magnitude of the
crossover length with the assumption that the crossover occurs in
the transition region of scaling of the mass with the dimension,
i.e., the first point where $p_+(l)$ and $p_-(l)$ become equal.\\
Then the crossover dimension is $l\sim \Delta$ for plates and
$l\sim S^{\frac{1}{2}}$ for rods, being $S$ the area of the basis
of the rod. This criterion, which is very acceptable physically,
is directly obtained in this formulation.\\ So, the usefulness of
Tsallis entropy to describe processes of fragmentation has been
tested in this work, where the FSDF for small and large fragments
have been obtained for fragmentation involving a change in the
geometry (dimensionality) of the fragments. We showed
analytically that the crossover detected in the experiments can
be obtained when the explicit scaling of the mass of the
fragments with the dimensionality is considered. In this respect
we have established that, as was pointed in \cite{1}, the slope is
determined by local rather than by global features of the
original object. Once again, the geometry is shown to be an important factor
in this phenomenon.\\ We point out again that non-extensive
statistics seem to have very much applications in non-equilibrium
phenomena, a number of which are yet to be investigated.\\

This work was partially supported by the ``Alma Mater'' contest, Havana University.
One of us (O.S) is grateful to the Department of Mathematical Sciences
 of Brunel University for
 kind hospitality and the Royal Society, London for financial support.

\newpage
\section{Table caption}
Table 1: Comparison of the results from \cite{1}with predictions
from \ref{eq:9}. Observe that all values of
$\frac{\beta_+}{\beta_-}$ lie in the predicted range
$\frac{2}{3}<\frac{\beta_+}{\beta_-}<1$.

\end{document}